\documentclass[11pt]{article}
\usepackage[utf8]{inputenc}
\usepackage{caption}
\usepackage{subcaption}
\usepackage{braket,slashed,bm}
\usepackage{jheppub}
\usepackage{simplewick}
\usepackage{array,multirow}
\usepackage{amsmath} 
\usepackage[normalem]{ulem}
\usepackage{calligra}
\usepackage{floatrow}
\DeclareUnicodeCharacter{2212}{\ensuremath{-}}
\DeclareMathAlphabet{\mathpzc}{OT1}{pzc}{m}{it}
\usepackage{mathrsfs}
\usepackage{tikz}
\usepackage{rotating}
\usepackage{pgf}

\usepackage{subcaption}
\usepackage{lscape}
\usepackage{graphicx}
\usepackage{booktabs}
\newfloatcommand{capbtabbox}{table}[][\FBwidth]

\newcommand{\lsim}{\mathrel{\mathop{\kern 0pt \rlap
  {\raise.2ex\hbox{$<$}}}
  \lower.9ex\hbox{\kern-.190em $\sim$}}}
\newcommand{\gsim}{\mathrel{\mathop{\kern 0pt \rlap
  {\raise.2ex\hbox{$>$}}}
  \lower.9ex\hbox{\kern-.190em $\sim$}}}

\DeclareUnicodeCharacter{2212}{\ensuremath{-}}

\allowdisplaybreaks
\begin{document}

\title{
Probing the Seed Effect of Supermassive Primordial Black Holes with Weak Gravitational Lensing 
}
\author[a,1]{Lei Zu,}
\author[a,b,c,d,e]{Kazunori Kohri}

\affiliation[a]{
Division of Science, National Astronomical Observatory of Japan (NAOJ),
2-21-1 Osawa, Mitaka, Tokyo 181-8588, Japan}

\affiliation[b]{
Department of Astronomy, The University of Tokyo,
Bunkyo-ku, Hongo, Tokyo 113-0033, Japan}

\affiliation[c]{
Department of Astronomical Science,
The Graduate University for Advanced Studies (SOKENDAI),
2-21-1 Osawa, Mitaka, Tokyo 181-8588, Japan}

\affiliation[d]{
Theory Center, IPNS, KEK,
1-1 Oho, Tsukuba, Ibaraki 305-0801, Japan}

\affiliation[e]{
Kavli IPMU (WPI), UTIAS, The University of Tokyo,
Kashiwa, Chiba 277-8583, Japan}

\emailAdd{lei.zu@astro.nao.ac.jp}

\note{East Asian Core Observatories Association (EACOA) Fellow}

\abstract{
We investigate the cosmological implications of supermassive primordial black holes (PBHs) as a component of dark matter (DM), focusing on their seed effect on large-scale structure formation. In contrast to the Poisson effect associated with the discrete distribution of PBHs, the seed effect arises because PBHs act as early gravitational seeds, accreting surrounding DM and developing dense DM spikes to form dressed PBHs. These structures enhance the matter power spectrum, particularly on small scales. We model their contribution within the halo-model framework under two prescriptions for DM accretion: an optimistic scenario (Case 1) and a conservative scenario (Case 2), which predict different modifications to the 1-halo term. We constrain the PBH abundance using weak-lensing measurements from the Dark Energy Survey Year 3 (DES Y3), performing a Markov Chain Monte Carlo analysis to derive upper limits on the PBH fraction $f_{\rm PBH}$, as a function of the PBH mass $M_{\rm PBH}$. We find that weak lensing places competitive constraints on supermassive PBHs, with the upper limits on $f_{\rm PBH}$ becoming progressively stronger toward higher PBH masses. Case 1, which predicts more efficient DM accretion and hence a larger enhancement of the matter power spectrum, yields stronger constraints than Case 2; in particular, it excludes $f_{\rm PBH}\gtrsim10^{-4}$ for $M_{\rm PBH}\gtrsim10^{10} M_\odot$. We further forecast the sensitivity of the Vera C. Rubin Observatory’s Legacy Survey of Space and Time (LSST) using mock weak-lensing data and find that the constraints could improve by approximately two orders of magnitude. Our results establish weak gravitational lensing as a sensitive probe of the seed effect of supermassive PBHs and demonstrate its potential to complement other cosmological and astrophysical searches.
}
\date{\today}
\maketitle
\section{Introduction}

The nature of dark matter (DM) remains one of the central unresolved questions in modern cosmology. Although its gravitational effects have been firmly established across a wide range of astrophysical and cosmological scales, its microscopic nature remains unknown. Primordial black holes (PBHs) provide a compelling non-particle DM candidate whose cosmological signatures can differ substantially from those of standard cold dark matter (CDM)~\cite{Hawking:1971ei,Carr:1974nx,Carr:1975qj,Kohri:2007qn,Khlopov:2008qy,Carr:2009jm,Frampton:2010sw,Kuhnel:2015vtw,Kawasaki:2016pql,Cotner:2016cvr,Bird:2016dcv,Georg:2017mqk,Ando:2017veq,Garcia-Bellido:2017aan,Inomata:2017okj,Kohri:2018qtx,Cotner:2018vug,Carr:2019hud,Carr:2020gox,Flores:2020drq,Kusenko:2020pcg,Green:2020jor,Villanueva-Domingo:2021spv}.

Recent observations with the James Webb Space Telescope (JWST) have identified candidate massive and luminous galaxies at high redshifts, including systems at $z > 10$, posing a potential challenge to the standard $\Lambda$CDM framework \cite{Boylan-Kolchin:2022kae,Labbe:2022ahb,Kobayashi:2026sji}. Supermassive PBHs have therefore been proposed as possible seeds that could accelerate the formation of massive high-redshift galaxies and alleviate this potential tension\cite{Liu:2022bvr,Yuan:2023bvh}. In addition, supermassive PBHs can contribute to additional sources for the
nano-hertz gravitational wave background reported by the pulsar-timing array observations~\cite{Kobayashi:2025jkg}. Such supermassive PBHs, if they exist, would affect not only the formation of individual early galaxies but also the distribution and clustering of matter~\footnote{If such massive PBHs were directly produced by collapses of large density perturbation, the non-detection of spectral distortions in the cosmic microwave background (CMB) excludes the masses larger
than $\sim 10^5 M_{\odot}$~\cite{Kohri:2014lza}. In this
paper therefore, we consider the other scenarios.}. Their strong gravitational potentials can alter the matter distribution, leaving signatures accessible to precision cosmological observables such as weak gravitational lensing.

The effects of PBHs on structure formation are commonly divided into two contributions: the Poisson effect and the seed effect~\cite{Bertschinger:1985pd,Afshordi:2003zb,Carr:2018rid,Inman:2019wvr,Carr:2026gwf}. The Poisson effect originates from fluctuations in the discrete number density of PBHs and produces a scale-independent contribution to the initial matter power spectrum. The seed effect, by contrast, arises because individual PBHs act as localized gravitational seeds that attract surrounding matter and accelerate the formation of nonlinear structures. Since a PBH represents a highly nonlinear initial overdensity rather than a small primordial density perturbation, matter can begin to accumulate around it considerably earlier than in the standard hierarchical formation scenario.

Previous large-scale-structure constraints on PBHs have primarily considered their Poisson contribution~\cite{Gong:2017sie,Murgia:2019duy,Mack:2006gz,Hu:2025adu,Ivanov:2025pbu}. However, the statistical description in terms of a Poisson power spectrum becomes insufficient on scales comparable to or smaller than the mean PBH separation, where the nonlinear structures surrounding individual PBHs must be treated explicitly. These scales are consequently excluded or effectively masked in many analyses~\cite{Ivanov:2025pbu,Liu:2022bvr}. For sufficiently massive and rare PBHs, corresponding to large $M_{\rm PBH}$ and small $f_{\rm PBH}$, the structures seeded by individual PBHs can provide an important, and potentially dominant, contribution. In this work, we focus on this seed contribution and adopt an idealized description in which the DM halos surrounding individual PBHs grow in isolation~\cite{Liu:2023pvq}.

When PBHs constitute only a subdominant fraction of the total DM abundance, they are embedded in an approximately smooth background of ambient DM~\cite{Bertone:2004pz}. Their gravitational potentials induce the accretion of surrounding DM, transforming initially bare PBHs into dressed PBHs surrounded by extended DM halos. Previous studies have shown that this process can generate steep and highly concentrated DM density profiles, often referred to as DM spikes~\cite{Oguri:2022fir,Mack:2006gz,Ricotti:2007au}. These dressed PBHs provide a physical realization of the seed effect: they behave as compact nonlinear structures whose masses and spatial profiles evolve through the accretion of ambient matter.

The population of dressed PBHs can substantially modify the matter power spectrum $P(k)$, particularly its one-halo contribution on small scales. In the standard halo model, the one-halo term describes correlations between matter elements residing within the same DM halo. Dressed PBHs introduce an additional population of unusually compact halos and can therefore enhance the power spectrum at large wavenumbers. Consequently, precision measurements of the matter distribution offer a promising avenue to detect these structures.

In this work, we incorporate dressed PBHs into the halo-model description of the nonlinear matter power spectrum. To account for uncertainties in the accretion history of the surrounding DM, we consider two prescriptions: an optimistic accretion scenario, denoted Case 1, and a conservative scenario, denoted Case 2. We then use cosmic-shear measurements from the DES Y3 to constrain the PBH fraction $f_{\rm PBH}$ as a function of the PBH mass $M_{\rm PBH}$. We additionally forecast the sensitivity of the Vera C. Rubin Observatory (LSST) using mock weak-lensing data. Our analysis demonstrates how weak gravitational lensing can probe the small-scale matter structures generated by the seed effect and provides constraints complementary to those derived from the conventional Poisson contribution.

\section{Dressed PBHs and the Matter Power Spectrum}

PBHs affect cosmic structure formation through mechanisms that differ from those associated with standard CDM. Because PBHs are discrete massive objects, fluctuations in their number density generate an irreducible Poisson contribution. For a monochromatic PBH population with comoving number density
\begin{equation}
\bar n_{\rm PBH}
=
\frac{f_{\rm PBH}\bar\rho_{\rm DM,0}}
{M_{\rm PBH}},
\end{equation}
the shot-noise power spectrum of the PBH number-density contrast is scale independent and given by
\begin{equation}
P_{\delta_{\rm PBH}}^{\rm shot}
=
\frac{1}{\bar n_{\rm PBH}},
\end{equation}
where $f_{\rm PBH}\equiv\Omega_{\rm PBH}/\Omega_{\rm DM}$ denotes the fraction of DM in PBHs~\cite{Gong:2017sie,Murgia:2019duy,Mack:2006gz,Hu:2025adu,Ivanov:2025pbu}.

The total matter power spectrum can thus be schematically represented as the sum of the standard adiabatic component and the PBH-induced isocurvature component:
\begin{equation}
    P(k,z) = P_{\rm ad}(k,z) + P_{\rm iso}(k,z).
\end{equation}
The adiabatic term $P_{\rm ad}(k,z)$ follows the standard $\Lambda$CDM evolution, while the isocurvature term is given by \cite{Mack:2006gz, Gong:2017sie}:
\begin{equation}
    P_{\rm iso}(k,z) = T^2_{\rm iso}(k) D^2(z) f^2_{\rm PBH} A_{\rm Poisson},
\end{equation}
where $D(z)$ is the linear growth factor, $f_{\rm PBH}$ is the fraction of DM composed of PBHs, and $T_{\rm iso}(k)$ is the isocurvature transfer function. Following \cite{Gong:2017sie}, the transfer function can be approximated as:
\begin{equation}
    T_{\rm{iso}}(k)= \left\{\begin{matrix} \frac{3}{2}(1+z_{eq})
  &  for\ k > k_{eq} \\
  0 & otherwise
\end{matrix}\right.,
\end{equation}
where $z_{\rm eq}$ and $k_{\rm eq}$ correspond to the redshift and scale of matter-radiation equality, respectively.

The linear Poisson description also becomes inadequate on sufficiently small scales. A useful characteristic wavenumber can be defined by requiring the dimensionless PBH shot-noise power,
\begin{equation}
\Delta_{\rm PBH}^{2}(k)
=
\frac{k^{3}}
{2\pi^{2}\bar n_{\rm PBH}},
\end{equation}
to reach unity. This gives
\begin{equation}
k_{\rm cut}
\equiv
\left(2\pi^{2}\bar n_{\rm PBH}\right)^{1/3}.
\end{equation}
Thus, $k_{\rm cut}$ characterizes the scale at which the discrete nature of the PBH population becomes nonlinear.

 Previous studies have extensively utilized this Poisson-induced enhancement to constrain PBH fractions through the Lyman-$\alpha$ forest\cite{Afshordi:2003zb, Murgia:2019duy, Ivanov:2025pbu}, weak lensing\cite{Hu:2025adu}, and 21-cm cosmology~\cite{Gong:2017sie}, the Poisson (or shot-noise) approximation has limitations. However, at $k\gtrsim k_{\rm cut}$, the description of the PBH population as a continuous linear density field is no longer sufficient. In this regime, the gravitational structures surrounding individual PBHs must be modeled explicitly. We are particularly interested in rare and massive PBHs, with
\begin{equation}
M_{\rm PBH}\gtrsim10^{8} M_\odot,
\qquad
f_{\rm PBH}\lesssim10^{-3}.
\end{equation}
Individual PBHs in this regime act as deep gravitational potential wells and accrete ambient DM, producing compact DM halos or spikes. This nonlinear "seed" contribution can exceed the extrapolated linear Poisson contribution over the small-scale range relevant to our analysis. We therefore do not extrapolate the linear Poisson power spectrum beyond $k_{\rm cut}$.~\cite{Bertschinger:1985pd,Villanueva-Domingo:2021spv}. 

In the idealized case where PBH-induced halos grow in isolation, the seed effect leads to the formation
of dressed PBHs, consisting of a central PBH surrounded by a compact halo of accreted DM.  During matter
domination, the gravitational potential of an individual PBH induces approximately radial infall of the
surrounding collisionless DM. Analytical calculations and numerical simulations suggest that the resulting
density profile can be approximated by a power law, $\rho(r) \propto r^{-\alpha}$ with $\alpha \approx 9/4$, extending to a characteristic accretion radius \cite{Berezinsky:2013fxa, Oguri:2022fir, Adamek:2019gns,Serpico:2020ehh,Boudaud:2021irr}. This accretion process is expected to remain efficient until $z \sim 30$, at which point the halo mass reaches $M_{\rm h} \sim 100 M_{\rm PBH}$, or more optimistically until the scale factor $a > a_{\rm crit} \equiv a_{\rm eq}/f_{\rm PBH}$ \cite{Liu:2023pvq}.

We incorporate the accreted DM structures into the halo-model framework by treating each PBH-seeded spike
as a localized matter perturbation. The 1-halo contribution to the matter power spectrum is defined as \cite{Cooray:2002dia}:
\begin{equation}
    P_{\rm 1h}(k) = \frac{1}{\bar{\rho}^2} \int \frac{dn}{dm} m^2 |u(k|m)|^2 dm,
\end{equation}
where $dn/dm$ is the halo mass function and $u(k|m)$ is the Fourier transform of the normalized density profile. For a monochromatic PBH mass distribution, the mass function simplifies to $n_{\rm PBH} \delta(M - M_{\rm h})$.

Because the matter contained in the spike is accreted from its surroundings, adding an uncompensated positive
halo profile would double-count part of the background matter density~\cite{Mack:2006gz}. We account for this redistribution by introducing a negative top-hat compensation shell extending from the halo radius $R_{\rm h}$ to a compensation radius $R_{\rm comp} = \alpha_{\rm comp} R_{\rm h}$ (setting $\alpha_{\rm comp}=5$ in this work). The modified Fourier-transformed profile is then:
\begin{equation}
    u(k) = \frac{1}{M_{\rm h}} \left[ \int_0^{R_{\rm h}} 4\pi r^2 \rho(r) \frac{\sin(kr)}{kr} dr - \int_{R_{\rm h}}^{R_{\rm comp}} 4\pi r^2 \rho_{\rm comp} \frac{\sin(kr)}{kr} dr \right],
\end{equation}
where $\rho_{\rm comp}$ is determined by mass conservation: $\rho_{\rm comp} = M_{\rm h} / [\frac{4}{3}\pi (R_{\rm comp}^3 - R_{\rm h}^3)]$. This ensures $u(k \to 0) \to 0$, consistent with the fact that the total mass fluctuation on large scales must vanish. 

To bracket the uncertainty in the late-time growth of PBH-seeded halos, we consider two limiting scenarios
for rare and massive PBHs with
$M_{\rm PBH} > 10^8 M_\odot$, $f_{\rm PBH} < 10^{-3}$: 
\begin{itemize}
    \item \textbf{Case 1 (Optimistic):} Accretion is assumed to persist until $a = a_{\rm crit}$. For the parameter space considered, this leads to $M_{\rm h} \gtrsim 1000 M_{\rm PBH}$, and we adopt $M_{\rm h} = 1000 M_{\rm PBH}$.
    \item \textbf{Case 2 (Conservative):} Accretion ceases at $z = 30$, yielding a halo mass $M_{\rm h} \sim 100 M_{\rm PBH}$.
\end{itemize}
In both cases, we assume that the masses and density profiles of the dressed-PBH halos remain unchanged after accretion ceases.

\begin{figure}[ht!]
\centering
\includegraphics[scale=0.6]{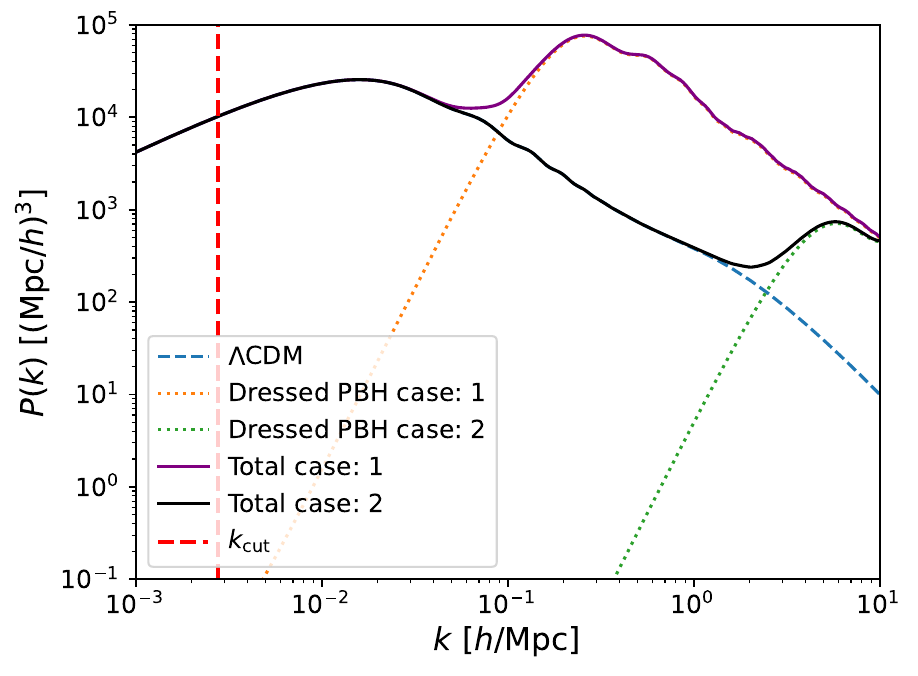}

\caption{
The matter power spectrum including the contribution of dressed-PBH halo. The figure displays the results for $M_{\rm PBH}=10^{15} M_{\odot}$ and $f_{\rm PBH}=10^{-5}$. The blue dashed line ($\Lambda$CDM) represents the non-linear matter power spectrum predicted by the standard $\Lambda$CDM model (calculated with HMCode~\cite{Mead_2016}). The orange dotted line (Dressed PBH case: 1) and green dotted line (Dressed PBH case: 2) illustrate the additional 1-halo term contributions for two distinct dressed PBH scenarios. The purple solid line (Total case: 1) and black solid line (Total case: 2) show the combined matter power spectra for each respective scenario ($P_{\rm total} = P_{\rm \Lambda \text{CDM}} + P_{\text{1h, Dressed PBH}}$). The vertical red dashed line ($k_{\rm cut}$) marks a characteristic wavenumber, $k_{\rm cut} = (2\pi^2 n_{\rm PBH})^{1/3}$, where $n_{\rm PBH}$ is the comoving number density of PBHs. At wavenumbers $k>k_{\rm cut}$, the Poisson effect's simple white noise contribution is often negligible compared to the dominant dressed PBH seed effect.
}
\title{Matter power spectrum at $z=0$ including the contribution from dressed PBHs.}
\label{mpk}
\end{figure}

To illustrate the impact of dressed PBHs on the matter distribution, we present the matter power spectrum, $P(k)$, at redshift $z=0$ for $M_{\rm PBH}=10^{15} M_{\odot}$ and $f_{\rm PBH}=10^{-5}$ in Fig.\ref{mpk}.  In this figure, the blue dashed line represents the standard non-linear $\Lambda$CDM matter power spectrum, computed using \texttt{HMCode}\cite{Mead_2016}. The orange and green dotted lines illustrate the additional 1-halo term contributions from the two dressed PBH scenarios (Case 1 and Case 2, respectively). Due to the mass compensation mechanism, the contribution from dressed PBHs is suppressed at large scales (small $k$), ensuring consistency with the background evolution. However, it rises sharply at smaller scales (large $k$), reflecting the extreme density and compactness of the DM spikes surrounding the PBHs. These results clearly demonstrate that dressed PBHs introduce a characteristic and potentially observable signature in the matter power spectrum at small, non-linear scales, offering a promising avenue to constrain PBH abundance and properties using LSS probes such as cosmic shear.

\section{Cosmic Shear}

To probe the modifications to the small-scale matter power spectrum induced by dressed PBHs, we use cosmic shear, one of the most sensitive observational probes of the projected matter distribution~\cite{DES:2017myr,DES:2021bvc,vanUitert:2017ieu,Heymans:2020gsg,HSC:2018mrq,Hamana:2019etx,LSST:2008ijt}. Cosmic shear refers to the weak distortions induced in the shapes of distant background galaxies by the gravitational lensing effect of intervening LSS. The accumulated deflection of light along the line of sight produces small but spatially coherent distortions in the observed galaxy shapes. Because gravitational lensing responds to the total gravitational potential, cosmic shear probes the distribution of matter without requiring it to emit light or to follow a prescribed galaxy-bias relation.

For a pair of tomographic source-redshift bins $i$ and $j$, the theoretical cosmic-shear correlation functions are given, in the flat-sky approximation, by
\begin{equation}
\xi_{\pm}^{ij}(\theta)
=
\frac{1}{2\pi}
\int_{0}^{\infty}
d\ell \  \ell J_{0/4}(\ell\theta)
C_{\rm tot}^{ij}(\ell),
\end{equation}
where $\ell$ is the angular multipole. The zeroth-order Bessel function $J_0$ applies to $\xi_+$, whereas the fourth-order Bessel function $J_4$ applies to $\xi_-$.

Under the extended Limber approximation~\cite{Limber:1953,LoVerde:2008re}, the gravitational shear-shear angular power spectrum between source bins $i$ and $j$ is related to the three-dimensional matter power spectrum $P_\delta(k,z)$ through
\begin{equation}
\label{eq}
C_{GG}^{ij}(\ell)
=
\int_{0}^{\chi_{\rm max}}
d\chi
\frac{W^i(\chi)W^j(\chi)}{\chi^2}
P_\delta
\left(
k=\frac{\ell+1/2}{\chi},
z(\chi)
\right),
\end{equation}
where $\chi$ is the comoving radial distance.

We model intrinsic alignments using the nonlinear alignment model(NLA)~\cite{Hirata:2004gc,Bridle:2007ft,Joachimi:2010xb}. The total angular power spectrum entering the shear correlation functions is therefore
\begin{equation}
C_{\rm tot}^{ij}(\ell)
=
C_{GG}^{ij}(\ell)
+
C_{GI}^{ij}(\ell)
+
C_{IG}^{ij}(\ell)
+
C_{II}^{ij}(\ell),
\end{equation}
where $C_{GI}^{ij}$ and $C_{IG}^{ij}$ describe the correlations between gravitational shear and intrinsic galaxy shapes, while $C_{II}^{ij}$ represents the intrinsic-shape correlation. 

For a spatially flat universe, the lensing-efficiency kernel for the $i$-th source bin is
\begin{equation}
W^i(\chi)
=
\frac{3H_0^2\Omega_{\rm m}}{2c^2}
\frac{\chi}{a(\chi)}
\int_{\chi}^{\chi_{\rm max}}
d\chi'
n^i(\chi')
\frac{\chi'-\chi}{\chi'},
\end{equation}
where ($H_0$) is the Hubble constant, $\Omega_{\rm m}$ is the present-day matter density parameter, $a(\chi)$ is the scale factor, and $c$ is the speed of light. The function $n^i(\chi)$ denotes the normalized comoving-distance distribution of source galaxies in the $i$-th tomographic bin.

\begin{figure}[ht!]
\centering
\includegraphics[scale=0.8]{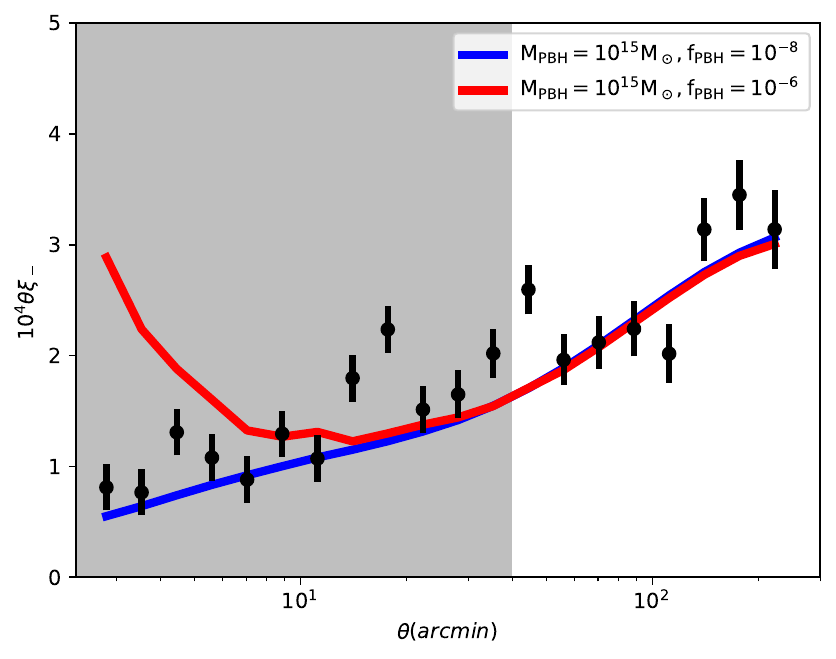}
\caption{Cosmic shear signal in the 4th-4th redshift bin($\xi_{- }$) of the DES Y3 data alongside predictions of the dressed-PBH halo. The gray region is masked to avoid the baryonic feedback uncertainty.}
\label{cosmic_shear}
\end{figure}

In this study, we utilize the DES Y3 data, which comprises shape measurements of over $10^8$ source galaxies covering an effective footprint of $A_{\rm{eff}} = 4143~\rm{deg}^2$~\cite{DES:2021bvc}. Following the standard DES Y3 methodology, the \texttt{METACALIBRATION} shape catalog is partitioned into four tomographic redshift bins across the range $0 < z < 3$~\cite{DES:2021bvc}. To mitigate uncertainties arising from baryonic effects, we apply a mask to the small angular scales following the criteria established in Ref.~\cite{DES:2021bvc}. We also utilize cosmic shear mock data from Vera C. Rubin Observatory(formerly known as the Large Synoptic Survey Telescope, LSST)~\cite{LSSTScience:2009jmu} for future forecasts. Fig.~\ref{cosmic_shear} illustrates the impact of dressed PBH halos on the cosmic shear signal. Specifically, it displays the predicted cosmic shear correlation function ($\xi_{-}$ in the 4th-4th redshift bin) for two representative PBH parameter sets: $M_{\rm PBH}=10^{15}M_\odot$ with $f_{\rm PBH}=10^{-8}$ (blue line) and $M_{\rm PBH}=10^{15}M_\odot$ with $f_{\rm PBH}=10^{-6}$ (red line), overlaid with the DES Y3 observational data. As expected, the scenario with a higher PBH fraction $f_{\rm PBH}=10^{-6}$ exhibits a more pronounced enhancement at small angular scales, reflecting the increased contribution from the dressed PBH 1-halo term.

To model the cosmological evolution, we employ the Boltzmann solver \texttt{CLASS}~\cite{Lesgourgues:2011re}, specifically adopting the \texttt{HMCode}~\cite{Mead_2016} prescription to compute the nonlinear matter power spectrum for the $\Lambda$CDM baseline. We then add the 1-halo contribution from dressed PBHs to this baseline. Global parameter estimation is performed using the Markov Chain Monte Carlo (MCMC) sampler \texttt{MontePython}~\cite{Audren:2012wb,Brinckmann:2018cvx}. 
In our analysis, we include the six $\Lambda$CDM parameters: the baryon $\Omega_{b}$ and CDM energy densities $\Omega_{\rm{CDM}}$, the amplitude $A_s$ and spectral index $n_s$ of primordial scalar perturbations, the optical depth to  reionization $\tau_{\rm{reio}}$, and the angular size of the horizon at the last scattering surface $\theta$. For the dressed PBHs component, we add two additional parameters: the mass of the PBH $\rm{M_{PBH}}$ and  the fraction of DM composed of PBHs $f_{\rm{PBH}}$.  The prior range of each parameters are shown in Table. \ref{tab:prior}.

\begin{table}
    \centering
    \begin{tabular}{|c|c|c|c|c|c|c|c|c|}
\hline
\hline
\hline
\multicolumn{2}{|c|}{\textbf{Cosmological Parameters}}\\
\hline
$100 \Omega_b h^2$  & [2.147, 2.327] \\
$\Omega_{\rm{CDM}} h^2$ &   [0 , 0.2] \\
$100 \theta_s$ &  [1.0393,  1.0429] \\
$\ln \left(10^{10} A_s\right)$  & [2.9547,  3.1347] \\
$n_s$ &   [0.9407, 0.9911]  \\
$\tau_{\text {reio }}$ &    [0.01,  0.7]  \\
$\log_{10}\frac{\rm{M_{PBH}}}{\rm{M_\odot}}$  & [8.0, 18.0] \\
$\log_{10}f_{\rm{PBH}}$  &  [-11, -3] \\
\hline
\multicolumn{2}{|c|}{\textbf{Intrinsic alignment}}\\
\hline
$A_{\rm IA}$ & [-6,6]\\
$\eta$ & [-6,6] \\
\hline
\hline

\end{tabular}
    \caption{The parameters and their prior ranges used in the MCMC analysis are listed above. We use flat priors within the specified ranges.}
    \label{tab:prior}
\end{table}

\section{Results}

\begin{figure}[ht!]
\centering
\includegraphics[scale=0.8]{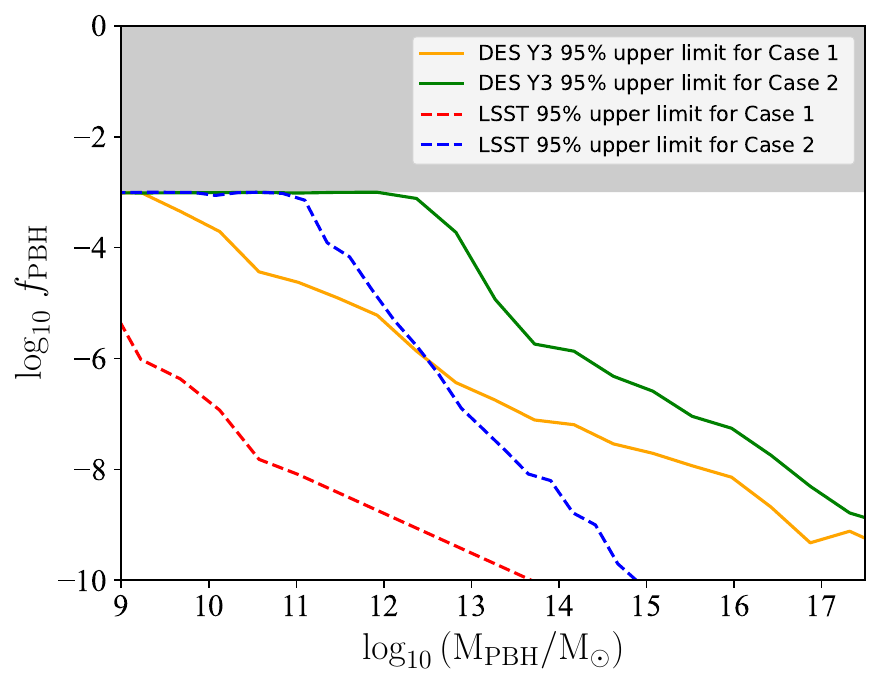}
\caption{The 95\% confidence level (C.L.) upper limits on the PBH fraction, $f_{\rm PBH}$, as a function of the PBH mass, $M_{\rm PBH}$. The constraints are derived from the DES Y3 cosmic shear data (solid lines) and the mock LSST cosmic shear data (dashed lines). For each dataset, two distinct scenarios for the DM spike evolution are shown, denoted as Case 1 and Case 2.}
\label{bounds}
\end{figure}

Figure~\ref{bounds} presents the 95\% confidence level (C.L.) upper limits on the PBH fraction, $f_{\rm PBH}$, as a function of the PBH mass, $M_{\rm PBH}$. We calculated these upper limits by using the profile likelihood method. The solid curves show the constraints obtained from the DES Y3 cosmic-shear data, while the dashed curves show the forecast sensitivities derived from the LSST-like mock data. We present results for the two dressed-PBH growth prescriptions introduced above: the optimistic accretion model (Case 1) and the conservative model (Case 2).

For both cases, the upper limit on $f_{\rm PBH}$ becomes progressively stronger with increasing $M_{\rm PBH}$. This behavior can be understood from the approximate scaling of the seed contribution: a more massive $M_{\rm PBH}$ seed facilitates the accretion of a larger DM halo ($M_h$), thereby shifting the characteristic scale of the seed effect toward larger physical scales. 

For Case 1, which represents a more aggressive accretion model, the DES Y3 data excludes $f_{\rm PBH}$ values above $10^{-4}$ for $M_{\rm PBH} \gtrsim 10^{10} M_\odot$. Case 2 yields relatively relaxed bounds compared to Case 1, particularly below $10^{12} M_\odot$, reflecting the impact of the halo mass accretion process on the 1-halo power enhancement. The LSST forecast improves the sensitivity by more than two orders of magnitude. This improvement results from the larger survey area, higher source-galaxy density, and enhanced statistical sensitivity of the LSST-like cosmic-shear data. Compared with existing constraints based primarily on PBH-induced Poisson isocurvature perturbations, including those from the Lyman-$\alpha$ forest and 21-cm cosmology~\cite{Afshordi:2003zb,Gong:2017sie,Murgia:2019duy,Ivanov:2025pbu}, our constraints are particularly competitive in the regime of rare and supermassive PBHs. Over the range of scales retained in our analysis, the nonlinear structures surrounding individual PBHs can generate a larger contribution than an extrapolation of the linear Poisson term. Weak gravitational lensing therefore provides a complementary probe of the PBH seed effect that is directly sensitive to the resulting redistribution of matter. 

These results do not constitute evidence for a dressed-PBH signal. Instead, they demonstrate that the absence of an observed small-scale excess in the cosmic-shear correlation functions places stringent limits on the abundance of supermassive PBHs. Future weak-lensing measurements from Rubin Observatory's LSST will substantially improve the sensitivity to this class of models.

\section{Discussion}

In this work, we have explored the cosmological implications of supermassive PBHs constituting a subdominant component of DM, focusing on their seed effect on the formation of LSS. PBHs gravitationally accrete the surrounding DM and develop compact DM spikes, thereby forming dressed PBHs. We incorporated these objects into a halo-model description of the nonlinear matter power spectrum and showed that their compensated one-halo contribution can enhance the matter power spectrum on small scales. Using cosmic shear data from the DES Y3, we derived 95\% confidence level (C.L.) upper limits on the PBH abundance under two prescriptions for the growth of the surrounding DM halos: an optimistic accretion scenario (Case 1) and a conservative scenario (Case 2). These results demonstrate that weak gravitational lensing, through its direct sensitivity to the projected distribution of gravitating matter, provides a powerful and complementary probe of rare and supermassive PBHs. We find no evidence for a dressed-PBH contribution in the DES Y3 data and therefore interpret our results as upper limits on $f_{\rm PBH}$.

Our analysis relies on several simplifying assumptions. In particular, we model each PBH-seeded halo as an isolated object and approximate its late-time evolution using two fixed accretion prescriptions. In a realistic cosmological environment, neighboring dressed halos may overlap, merge into larger host halos, or undergo tidal stripping. These processes can alter both the mass and internal density profile of the DM spikes, potentially redistributing or suppressing their contribution to the small-scale matter power spectrum. The phenomenological compensation profile adopted in this work also provides only an approximate description of the matter removed from the surroundings during halo growth.

Additional effects may become important in the regime of extremely massive and rare PBHs. The finite number of PBHs within the survey volume can generate significant non-Gaussian covariance and sample variance, while correlations between PBH-seeded halos and the standard large-scale matter distribution may invalidate the assumption that their power-spectrum contribution can be added independently to the $\Lambda$CDM prediction. Consequently, the constraints at the highest PBH masses should be interpreted within the idealized halo-model framework adopted here.

A more complete treatment will require high-resolution cosmological simulations capable of following the formation, overlap, tidal evolution, and hierarchical merging of PBH-seeded halos. Such simulations would allow the density profiles, compensation scales, halo occupation statistics, and non-Gaussian covariance to be modeled self-consistently. They would also clarify the range of scales over which the isolated seed-effect description remains reliable.

Forthcoming weak-lensing measurements from Rubin Observatory's Legacy Survey of Space and Time and from Euclid will substantially improve the statistical sensitivity to dressed PBHs. Joint analyses combining cosmic shear with galaxy clustering, CMB lensing, the Lyman-$\alpha$ forest, and 21-cm observations will provide complementary scale and redshift information, help break parameter degeneracies, and enable more robust tests of the remaining PBH parameter space.

\begin{acknowledgments}
This work was in part supported by the JSPS KAKENHI Grant No. JP24K07027 (KK). L.Z acknowledges support from the East Asian
Core Observatories Association Fellowship. 
\end{acknowledgments}

\bibliographystyle{JHEP}
\bibliography{ref.bib}

\appendix

\end{document}